\documentclass[conference]{IEEEtran}

\usepackage{graphicx}
\usepackage{url,amstext,amsfonts,amssymb,amscd,amsbsy,amsmath,amsthm,verbatim,mathrsfs}
\usepackage{ifthen, fullpage, mathrsfs, cite,hyperref,framed}
\usepackage{color}
\usepackage{algorithm}
\usepackage{algpseudocode}
\usepackage{longtable}
\usepackage{rotating}

\usepackage{caption}
\DeclareCaptionFormat{myformat}{#3}

\newtheorem{theorem}{Theorem}[section]

\newtheorem{prop}[theorem]{Proposition}

\theoremstyle{definition}

\theoremstyle{remark}

\begin{document}

\title{Geolocation with FDOA Measurements via Polynomial Systems and RANSAC}

\author{\IEEEauthorblockN{Karleigh J. Cameron}
\IEEEauthorblockA{Dept. of Mathematics\\
Colorado State University\\
Fort Collins, C0, 80523 (USA)\\
Email: cameron@math.colostate.edu}
\and
\IEEEauthorblockN{Daniel J. Bates}
\IEEEauthorblockA{Dept. of Mathematics\\
Colorado State University\\
Fort Collins, C0, 80523 (USA)\\
Email: bates@math.colostate.edu}} 

\maketitle

\begin{abstract}
The problem of geolocation of a transmitter via time difference of arrival (TDOA) and frequency difference of arrival (FDOA) is
given as a system of polynomial equations.  This allows for the use of homotopy continuation-based methods from numerical algebraic geometry. A novel geolocation algorithm employs numerical algebraic geometry techniques in conjunction with the random sample consensus (RANSAC) method. This is all developed and demonstrated in the setting of only FDOA measurements, without loss of generality. Additionally, the problem formulation as polynomial systems immediately provides lower bounds on the number of receivers or measurements required for the solution set to consist of only isolated points.
\end{abstract}

\maketitle


\section*{Introduction}

The determination of the location of an RF transmitter based on measurements from several receivers is a fundamental problem in
a number of applications.  These measurements include altitude (ALT), time difference of arrival (TDOA), and frequency difference of arrival (FDOA) ~\cite{Ho1997}.  In fact, all of these measurements can be cast as polynomials
in the variables corresponding to the location of the transmitter.  The problem of geolocation can then be solved by computing numerical approximations
of the solutions of polynomial systems, and it is simple to immediately provide lower bounds on the number of measurements needed to reduce the
solution set dimension to zero (i.e., so there are only finitely many potential locations). \\

Given a system of polynomial equations, the mathematical area called numerical algebraic geometry provides a range of tools and software for the solution of polynomial systems. Pairing this set of tools with the random sample consensus (RANSAC) method then yields a novel approach to RF transmitter geolocation.   \\

The case of solving for emitter location using only TDOA measurements  is simpler and has been well studied. Ho and Chan~\cite{Ho1997}
provided a first development of equations for using TDOA  measurements to back out emitter location, coming very close to the polynomial
system development given in the next section. The more recent articles~\cite{Compagnoni2014,Compagnoni2016} provide an advanced
description of this TDOA-only setting, based largely on algebraic geometry.  Various other approaches have been developed, and at least
one~\cite{Shuster} has made use of numerical algebraic geometry, though in a rather different context.  RANSAC has also previously been
applied to the TDOA-only setting~\cite{Li2009}. \\

Although the methods developed in this paper can be used to locate an emitter using TDOA measurements only, FDOA measurements only, or a combination of both, we choose to focus on the FDOA-only case. Doppler resolution is higher than range resolution for signals with narrow-bandwidth and long pulse duration~\cite{Cheney2009,Mason2005}, making it desirable to solve for emitter location using FDOA alone in these cases. Source localization in the FDOA-only case, however, is far more nuanced, as the geometry associated with the FDOA measurements is more interesting and complicated~\cite{Cheney2017}. \\

For geolocation using FDOA alone in particular, there are often multiple possible emitter locations corresponding to observed measurements. This can cause problems for iterative methods that converge to a single solution~\cite{Mason2005}. In contrast, numerical algebraic geometry techniques will find all real, feasible solutions. Our method then uses RANSAC to help determine which of these solutions is most consistent with other data gathered.

In~\S\ref{s:Background} we formulate the TDOA- and FDOA-based source localization problem as polynomial systems and provide an introduction to methods from numerical algebraic geometry to be used for their solution. The proposed geolocation algorithm (FDOAR) incorporating numerical algebraic geometry techniques and the iterative process, RANSAC, is presented in~\S\ref{s:RANSAC} along with numerical results.  A novel upper bound on feasible FDOA measurements for use in denoising data is presented in~\S\ref{s:FDOA_bounds}.  In~\S\ref{s:Bounds} we present lower bounds on the number of receivers or measurements needed to reduce the solution set to only isolated points. Benefits and limitations of the FDOAR geolocation algorithm are discussed in~\S\ref{s:Disc}.

The primary contributions of this article are
\begin{itemize}
\item a development of the geolocation problem via TDOA and FDOA measurements as polynomial systems;
\item a novel upper bound on possible FDOA measurements; and
\item lower bounds on the number of receivers or measurements needed to reduce the solution set to only points;
\item a novel approach to geolocation using the pairing of numerical algebraic geometry with RANSAC.
\end{itemize}


\section{Background}\label{s:Background}

\subsection{Polynomial Systems for Geolocation}\label{s:GeoPoly}

Using a similar setup to that in ~\cite{Ho1997}, the relationship between TDOA measurements, FDOA measurements, and transmitter location can be represented as a set of polynomials. Consider a system of receivers, labeled $1,\hdots,N.$ Without loss of generality, the first receiver can be chosen as a reference receiver, such that TDOA ($\tau_{i,1}$) and FDOA ($f_{i,1}$) can be calculated between receivers $i$ and 1, where $i=2,\hdots,N$, for a total of $N-1$ pairs of measurements. The problem can be cast as two- or three- dimensional. For generality, we choose to use a three-dimensional earth-centered, earth fixed (ECEF) coordinate system. Thus, each receiver has known location $\mathbf{x}_i=[x_i,y_i,z_i]^T$ and velocity $\dot{\mathbf{x}}_i=[\dot{x}_i,\dot{y}_i,\dot{z}_i]^T.$ It is desirable to solve for the location of a radio-frequency emitter, $\mathbf{x}=[x,y,z]^T$.

\subsubsection{Time Difference of Arrival (TDOA)}

The amount of time it takes for the signal to travel from $\mathbf{x}$ to $\mathbf{x}_i$ is $\|\mathbf{x}-\mathbf{x}_i\|/c$, where $c$ is the speed of propagation. Thus the TDOA between receiver $i$ and receiver 1 is equivalent to
\begin{align*}
c\cdot\tau_{i,1} &= \|\mathbf{x}_i-\mathbf{x}\| - \|\mathbf{x}_1-\mathbf{x}\| \\
 &= \sqrt{(x_i-x)^2+(y_i-y)^2+(z_i-z)^2} \\
 & \qquad - \sqrt{(x_1-x)^2+(y_1-y)^2+(z_1-z)^2}.
\end{align*}
\noindent Considering all $N$ receivers, this system can be transformed into the set of polynomial equations~\cite{Ho1997}:
\begin{flalign*}
&(c\cdot \tau_{1,2})^2 + 2cr_1 \cdot \tau_{1,2}\\
& - (x_2^2 + y_2^2 + z_2^2) + (x_1^2 + y_1^2 + z_1^2) \\
& + 2\left[(x_2-x_1)x + (y_2-y_1)y + (z_2-z_1)z\right] = 0 \\ \\
&(c\cdot \tau_{1,3})^2 + 2cr_1 \cdot \tau_{1,3} \\
&- (x_3^2 + y_3^2 + z_3^2) + (x_1^2 + y_1^2 + z_1^2) \\
& + 2\left[(x_3-x_1)x + (y_3-y_1)y + (z_3-z_1)z\right] = 0 \\
& \qquad \qquad\qquad \qquad \vdots \\
&(c\cdot \tau_{1,N})^2 + 2cr_1 \cdot \tau_{1,N}\\
&- (x_N^2 + y_N^2 + z_N^2) + (x_1^2 + y_1^2 + z_1^2) \\
& + 2\left[(x_N-x_1)x + (y_N-y_1)y + (z_N-z_1)z\right] = 0 \\ \\
& r_1^2 - (x^2+y^2+z^2) - (x_1^2+y_1^2+z_1^2)\\
& \qquad \qquad \qquad \qquad \;+2 (x_1x + y_1y +z_1z) = 0,
\end{flalign*}
where variable $r_1$, representing the range of receiver 1, is used to remove square roots from the system.

\subsubsection{Frequency Difference of Arrival (FDOA)}\label{s:FDOA}

As developed in~\cite{Ho1997}, the FDOA between receiver $i$ and receiver 1 is,
\begin{align}
  \label{FDOA1}
f_{i,1}= \frac{f_0}{c}\left[\dfrac{\dot{\mathbf{x}}_i^T(\mathbf{x}_i-\mathbf{x})}{\|\mathbf{x}_i-\mathbf{x}\|}-\dfrac{\dot{\mathbf{x}}_1^T(\mathbf{x}_1-\mathbf{x})}{\|\mathbf{x}_1-\mathbf{x}\|}\right],
\end{align}
where $f_0$ is the emitted frequency. Although more complicated than the TDOA case, the equations above can be converted to a polynomial system with the addition of more range variables, $r_i$. The system becomes,
\begin{align*}
  \tiny
  &r_1 r_2 f_{1,2} \; \\
  & - \; r_1\left[\dot{x}_2(x_2-x) + \dot{y}_2(y_2-y) + \dot{z}_2(z_2-z)\right] \\
  &+ \; r_2\left[\dot{x}_1(x_1-x) + \dot{y}_1(y_1-y) + \dot{z}_1(z_1-z)\right] =0 \\
  &r_1 r_3 f_{1,3} \; \\
  & - \; r_1\left[\dot{x}_3(x_3-x) + \dot{y}_3(y_3-y) + \dot{z}_3(z_3-z)\right] \\
  &+ \; r_3\left[\dot{x}_1(x_1-x) + \dot{y}_1(y_1-y) + \dot{z}_1(z_1-z)\right] =0 \\
  &\qquad \qquad \qquad\qquad \; \; \vdots \\
   &r_1 r_N f_{1,N} \; \\
   &- \; r_1\left[\dot{x}_N(x_N-x) + \dot{y}_N(y_N-y) + \dot{z}_N(z_N-z)\right] \\ &+ \; r_N\left[\dot{x}_1(x_1-x) + \dot{y}_1(y_1-y) + \dot{z}_1(z_1-z)\right] =0 \\ \\
   &r_1^2 - (x^2+y^2+z^2) - (x_1^2+y_1^2+z_1^2) \\
   &\qquad \qquad \qquad \qquad \;+ 2 (x_1x + y_1y +z_1z) = 0 \\
  &\qquad \qquad \qquad \qquad \; \;\vdots \\
  &r_N^2 - (x^2+y^2+z^2) - (x_N^2+y_N^2+z_N^2) \\
  &\qquad \qquad \qquad \qquad \;+ 2 (x_Nx + y_Ny +z_Nz) = 0.
\end{align*}


\subsection{Numerical Methods for Polynomial Systems}\label{s:NAG}

Rephrasing the above problems as systems of polynomial equations now opens the door to methods from algebraic geometry,
particularly numerical algebraic geometry.  The core computational engine of this field is homotopy continuation.  The idea is as
follows:  Given a polynomial system of equations $\mathbf{f}(\mathbf{z}) = 0$ to be solved ($\mathbf{z}\in\mathbb C^N$), a related system
$\mathbf{g}(\mathbf{z})$ is constructed.  For example, the {\em total degree start system} for a system $\mathbf{f}(\mathbf{z})$
is given by $g_i(\mathbf{z}) = z_i^{d_i}-1$, where $d_i$ is the degree of the polynomial $f_i(\mathbf{z})$.  A homotopy function,
typically
$$\mathbf{H}(\mathbf{z},t) = t\ \mathbf{g}(\mathbf{z}) + (1-t)\ \mathbf{f}(\mathbf{z}),$$
between these two systems is then constructed so that $\mathbf{H}(\mathbf{z},1) = \mathbf{g}(\mathbf{z})$ and
$\mathbf{H}(\mathbf{z},0) = \mathbf{f}(\mathbf{z})$.  Some basic algebraic geometry then guarantees that each isolated solution
of $\mathbf{f}(\mathbf{z})=0$ will be reached by at least one path that varies in $t$ and includes a solution of $\mathbf{g}(\mathbf{z})=0$.
Thus, to find all isolated solutions of $\mathbf{f}(\mathbf{z})=0$, it suffices to find all solutions of $\mathbf{g}(\mathbf{z})$ and use
numerical path-tracking methods~\cite{ag90} to move from $t=1$ to $t=0$.  Predictor-corrector methods are the standard choice.
Extraneous paths will diverge, but this wasted computation time can be partially mitigated by terminating any path with norm above
some threshold or working in projective space instead of affine space.


This is only a very basic description of a procedure that is fundamental in dozens of methods.
For example, there are numerical methods for finding all complex positive-dimensional solution components (curves, surfaces, etc.),
for extracting real solutions in various circumstances, for making use of particular polynomial system structure to reduce run time,
and so on.  For many further details, see~\cite{SW05,BertiniBook} and the references therein.  For now,
it suffices to know a bit more about singular solutions, ill-conditioning, and parameter homotopies.

\subsubsection{Singularity}
Just as $(x-1)^2=0$ has a double root at $x=1$, isolated solutions of polynomial systems can have multiplicity greater than 1.
The Jacobian matrix (the matrix of all first partial derivatives of the system) is singular at such solutions, so these solutions are referred
to as singular.  The basic methods of numerical algebraic geometry include so-called {\em endgames}, specialized techniques based on
Puiseux series expansions or the Cauchy integral formula, which can be used to accurately compute numerical approximations to
singular solutions of $\mathbf{f}(\mathbf{z})=0$.  However, when tracking paths before $t=0$, paths can become very
close\footnote{Actual path crossing is a probability 0 event~\cite{SW05,BertiniBook}.}, causing the Jacobian matrix to become
ill-conditioned and resulting in poor path-tracking performance.  This is mitigated in the software package Bertini~\cite{Bertini} via
step size control and adaptive precision techniques, but can still lead to various recognizable tracking failures~\cite{BertiniBook}.

\subsubsection{Parameter Homotopy}
Finally, in the special setting of repeatedly solving polynomial systems with the same monomial structure but varying coefficients,
there is a particular valuable tool called the {\em parameter homotopy}.  The idea is as follows:  Suppose we wish to find the solutions of a parameterized
polynomial system of equations, $\mathbf{f}({\mathbf z};\mathbf{p})=0$, at each of a large number $n$ of parameter values
$\mathbf{p}_i\in\mathcal P$, $i=1,\ldots,n$ within some Euclidean (or possibly more general) parameter space $\mathcal P$.  We could
use the basic homotopy continuation mechanism described above, but there is a chance that much computational time will be wasted
tracking (possibly many) divergent paths for each $\mathbf{p}_i$.  Instead, the idea of the parameter homotopy is to first solve
$\mathbf{f}(\mathbf{z};\mathbf{p_0})$ for some random complex $\mathbf{p}_0\in\mathcal P$, then start from only the finite solutions at
$\mathbf{p}_0$ to track to each $\mathbf{p}_i$.  Once again, some basic algebraic geometry guarantees with probability 1 that the
number of finite, isolated solutions at $\mathbf{p}=\mathbf{p}_0$ will be the maximum number of finite, isolated solutions at any choice of $\mathbf{p}$.





\section{Numerical Algebraic Geometry and RANSAC for FDOA-based Geolocation}\label{s:RANSAC}

With a noiseless system, the numerical algebraic geometry methods above can be used to find all solutions to the FDOA system presented in~\S\ref{s:FDOA} accurate to any prescribed numerical accuracy. Specifically, it would take only a single solve in a software such as Bertini~\cite{Bertini} to obtain an emitter location. However, there are a couple issues that arise in real world situations. First, noise and measurement error can plague FDOA calculations and receiver location and velocity estimates. Additionally, if the receivers are positioned in a singular configuration or near one, computing the solution may be prohibitively expensive and the solution itself could be much more accurate in some coordinates than in others.  The nonlinear nature of the problem implies that there will often be multiple real solutions, which translate to multiple potential emitter locations. A robust accompaniment for the numerical algebraic geometry methods above is an iterative process such as the RANdom SAmple Consensus (RANSAC) algorithm.

RANSAC, originally developed~\cite{ransac} for application to the location determination problem, is useful when one has data with outliers or corrupt data points. The algorithm works by choosing a few samples from a set of data, determining a model to fit the samples, then calculating how many of the remaining data points can be considered inliers with respect to that model, up to a predetermined tolerance. This process is repeated for a prescribed number of iterations, then the model with the most inliers is returned.

As noted above, using RANSAC for geolocation is not a new idea. In fact, Li et al. applied the algorithm to source location with TDOA in~\cite{Li2009}. This paper proposes a modification of RANSAC to solve for source location with the FDOA polynomial system, a problem that is now accessible due to the utilization of numerical algebraic geometry techniques.

The most notable benefit of using RANSAC for this problem is the ability to ``ignore" noisy or corrupt data. Additionally, since many FDOA measurements are needed for the algorithm, it is natural to reformulate the polynomial system presented in~\S\ref{s:FDOA} to allow for measurements to be taken over multiple time steps. This reduces the number of  receivers needed to a single pair, with each system composed of FDOA measurements from three separate time steps (see~\S\ref{s:Bounds} for why data from three time steps is needed).

The algorithm outlined below involves solving a system using the numerical algebraic geometry software, Bertini \cite{Bertini}, during each iteration. Since each system will be of the same form and change only in certain parameter values (location, velocity, FDOA measurements), the solve can be structured as a parameter homotopy~\cite{SW05,BertiniBook}. As discussed in \S\ref{s:NAG}, this allows for only necessary paths to be tracked, which provides faster run times. Additionally, when the solving of an FDOA system results in multiple real, feasible solutions, we have modified the algorithm to consider each solution separately. This ensures there are no missed solutions, as can often result from iterative geolocation methods that converge to a single solution~\cite{Mason2005}.

\subsection{FDOA-RANSAC (FDOAR) Algorithm Outline\\}
\begin{framed}
\noindent{\bf Input}:
\begin{itemize}
  \item Locations, velocities, and corresponding FDOA measurement ($f_i$) for $n$ pairs of receivers (or $n$ timesteps of 1 pair of receivers).
  \item Number of iterations to run algorithm ($maxiter$).
  \item Inlier tolerance ($\varepsilon$).
\end{itemize}

\noindent{\bf Output}:
\begin{itemize}
  \item Estimated transmitter location, $\mathbf{x}$.
\end{itemize}

\begin{enumerate}
  \item Select three sample points from receiver data (FDOA measurements, receiver locations and velocities).
  \item Solve for emitter location using Bertini.
  \item Determine feasible solutions (must have positive range values and satisfy Prop. \ref{FDOAbound}).
  \item For each feasible solution, determine the number of sample pairs that can be considered inliers.
  \begin{enumerate}
    \item For each pair of receivers, determine the theoretical FDOA measurement, $\hat{f}_i$, corresponding to the solution.
    \begin{itemize}
      \item If $|\hat{f}_i-f_i|<\varepsilon$, mark sample as an inlier.
    \end{itemize}
    \item If the number of inliers for the current solution is greater than the previous recorded location, record current solution as best source location estimate.
  \end{enumerate}
  \item Repeat for designated number of iterations and return transmitter location estimate.
\end{enumerate}
\end{framed}

\subsection{Numerical Performance}\label{ss:Results}

\begin{figure}
  \centering
\includegraphics[scale=0.33]{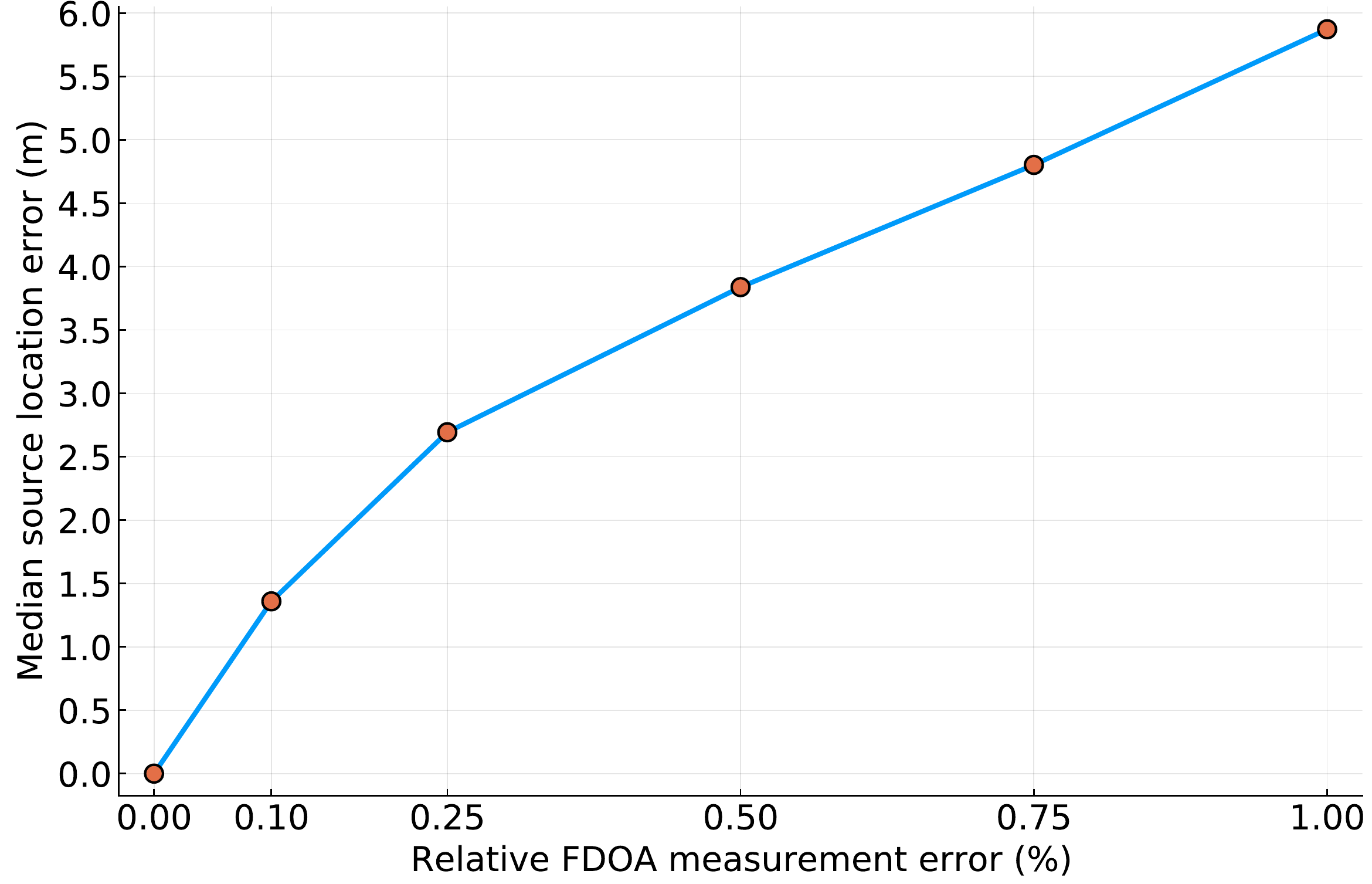}
\caption{Error in emitter location resulting from various levels of FDOA measurement error. For each data point, 50 instances of coupled RANSAC and Bertini were run, each having 20 iterations and $\varepsilon=0.03$.}
\label{numerr}
\end{figure}

Numerical simulations were run as follows. Consider a
Cartesian cube of space, 100m long on each side. For each numerical trial, a transmitter was placed at a random location, $\mathbf{x}$, in the cube. Locations and velocities for 40 pairs of receivers were also generated, with locations being limited to the interior of the cube and velocities in the range $[-2,2]$ m/s in each (x,y,z) direction. This is meant to simulate 40 time steps for a single pair of receivers and a stationary transmitter. For each pair of receivers, the FDOA was calculated according to Eq. \ref{FDOA1} and noise was added to simulate various levels of relative FDOA measurement error. We define this,
\begin{align*}
\small
\text{Relative FDOA Measurement Error }:=\dfrac{\sigma^2_{noise}}{\sigma^2_{FDOA}}\times 100\%,
\end{align*}
where $\sigma^2_{noise}$ and $\sigma^2_{FDOA}$ are the variance of the noise and variance of observed FDOA, respectively. The FDOAR algorithm was then run for 20 iterations, returning final transmitter estimate $\hat{\mathbf{x}}.$ The error for the trial was then calculated: $\|\hat{\mathbf{x}}-\mathbf{x}\|$ (m). Results are shown in Fig. \ref{numerr}. For each data point, this process was repeated 50 times and  the median of the error was recorded.

Many of the worst performing trials above resulted from transmitters located near the edges of the Cartesian box. We hypothesize that this is the result of very few (or none) of the receiver pairs being located on the side of the transmitter closest to the edge of the box. This caused less information to be learned about the transmitter and resulted in a worse estimate. This is consistent with geolocation intuition and suggests that error values in Fig. \ref{numerr} would decrease if one could ensure that receivers view the emitter from a variety of angles.

\section{Note on Denoising}\label{s:FDOA_bounds}

Since one of the key contributions of this article is the use of RANSAC for denoising data, we
include here a brief result that allows us to immediately remove FDOA
measurements that are physically unrealizable due to measurement error or noise.

\begin{prop} \label{FDOAbound} The frequency difference of arrival between receivers $i$ and $j$, $f_{i,j}$ satisfies:
\begin{equation*}
  \left|f_{i,j}\right|\leq \|\mathbf{v}_j\|+\|\mathbf{v}_i\|,
\end{equation*}
where $\mathbf{v}_i$ and $\mathbf{v}_j$ are the velocity vectors of receivers $i$ and $j$, respectively.
\end{prop}
\begin{IEEEproof}
\begin{align*}
&\left|f_{i,j}\right| = \left|\dfrac{\mathbf{v}_j\cdot (\mathbf{x}_j-\mathbf{x})}{\|{\mathbf{x}_j-\mathbf{x}}\|} -\dfrac{\mathbf{v}_i\cdot (\mathbf{x}_i-\mathbf{x})}{\|{\mathbf{x}_i-\mathbf{x}}\|}\right| \\[7pt]
&= \left|\dfrac{\|\mathbf{v}_j\|\|\mathbf{x}_j-\mathbf{x}\|\cos (\theta_j)}{\|{\mathbf{x}_j-\mathbf{x}}\|} -\dfrac{\|\mathbf{v}_i\|\|\mathbf{x}_i-\mathbf{x}\|\cos (\theta_i)}{\|{\mathbf{x}_i-\mathbf{x}}\|}\right| \\[7pt]
\vspace{2in}
&= \left|\|\mathbf{v}_j\|\cos (\theta_j)-\|\mathbf{v}_i\|\cos (\theta_i)\right| \\[7pt]
&\leq \|\mathbf{v}_j\|+\|\mathbf{v}_i\|.
\end{align*}
\end{IEEEproof}

It would also be interesting to consider denoising via projection to the manifold of
realizable FDOA measurements, similar to the use of projection in linear regression.
A similar approach was previously taken in the TDOA case~\cite{Compagnoni2017}.
We leave this for future work.



\section{Bounds on the Necessary Number of Measurements}\label{s:Bounds}


For systems of linear equations, it is trivial to predict the dimension of the solution set under the assumption
that the equations are linearly independent.  This is much the same with polynomial systems, though the
range of degenerate cases is far more nuanced and complicated.  With the formulation of the geolocation problem
as a system of polynomial equations in~\S\ref{s:GeoPoly}, it is easy to provide bounds on the minimum number of TDOA and FDOA
measurements\footnote{If we do not allow receivers to take measurements over multiple time steps, a similar table could be provided showing bounds on the number of receivers necessary. Here we refer to the number of measurements rather than the number of receivers for generality.}
needed in various scenarios to reduce the solution set to a finite set of points. The case is also considered where altitude of the emitter is known (ALT constraint). This is the content of Table 1.

\begin{center}
\begin{table}
\begin{tabular}{c|c|c}
& \# measurements (2D) & \# measurements (3D) \\
\hline
TDOA only & 2 & 3 \\
TDOA + ALT & - & 2 \\ \hline
FDOA only & 2 & 3 \\
FDOA + ALT & - & 2\\ \hline
TDOA + FDOA & 1 & 2 \\
TDOA + FDOA + ALT & - & 1
\end{tabular}
\caption{Minimum number of TDOA and FDOA measurements necessary to reduce set of potential transmitter locations to a finite number, for varying dimensions (2 or 3) and types of measurements being used.}
\end{table}
\end{center}

It is important to note that these bounds do not guarantee that there will be only finitely many solutions for every
set of measurements.  As an extreme counterexample, consider the case of stacking all receivers at the same
point; the number of (identical) measurements in this case makes no difference.

It is also worth noting that an anomalous positive-dimensional component (with $r_1=0$) shows up in the
FDOA only case.  However, this component is easily ignored as it is not physically feasible.


\section{Discussion}\label{s:Disc}

\subsection{Benefits of FDOAR}

We summarize a few of the primary benefits of our approach here:

\begin{enumerate}
\item Solving the geolocation systems using numerical algebraic geometry techniques finds all possible emitter locations. Coupling with RANSAC provides a way to determine which one of those locations best matches the rest of the data.
\item Any bad data from path failures, inaccuracies, measurement error, etc. is automatically ignored, assuming the
source of the errors is not implicit in the structure of the problem.
\item Our method uses FDOA measurements only, though it can be adapted to other measurement combinations.
\item Using multiple time steps, it is necessary to use only two receivers. Additionally, there is no need to designate a reference receiver, which could corrupt all data points if there are errors in its location or velocity.
\item When performing polynomial system solves at multiple points in parameter space, parameter
homotopies could improve efficiency.
\end{enumerate}

\subsection{Limitations}

Each path tracked when solving a polynomial system requires dozens, sometimes hundreds, of linear solves.  As a result,
any polynomial systems approach will necessarily be slower than any linear approach.  However, linearization necessarily
introduces inaccuracy to nonlinear problems, so the trade-off between speed and accuracy might lead different users to
use different approaches.

As with any RANSAC implementation, speed and accuracy is in part dependent upon the users choice of the maximum number of iterations and inlier tolerance. The optimal choice for these variables can depend greatly on the specifics of the problem. Theoretical results exist that bound the maximum number of iterations with respect to the percentage of inliers present in the data~\cite{Urbancic2014}.

\section{Future Work}

The problem of transmitter geolocation is mathematically rich and practically valuable.  As a result, there are many potential
avenues worthy of consideration.

Given a configuration of receivers and a generic set of measurements, it should be possible to decompose the space of
emitter locations into chambers corresponding to the number of physically realizable solutions of the corresponding
geolocation polynomial system.  An analysis over some set of such configurations could then help in choosing ``good''
receiver configurations.  Similarly, methods such as gradient descent homotopies~\cite{GDHom} could be useful
in finding the boundaries (called the {\em discriminant locus}) between these chambers.

As described above, it would be interesting to understand and make use of the semialgebraic set of physically
realizable FDOA measurements (see~\S\ref{s:FDOA_bounds}).  Perhaps this would provide some intuition for the estimation of geolocation accuracy.


\section*{Acknowledgments}

Both authors were partially supported by NSF grant DMS--1719658 and AFOSR grant FA9550-14-1-0185.  The authors wish to thank the PI of that AFOSR grant, Margaret Cheney, for introducing us to this problem and numerous stimulating discussions.  We also wish to thank Jon Hauenstein for suggesting the use of RANSAC.

\bibliographystyle{abbrv}

\end{document}